\documentclass[graybox]{svmult}

\usepackage{type1cm}        
\usepackage{makeidx}         
\usepackage{graphicx}        
\usepackage{multicol}        
\usepackage[bottom]{footmisc}
\usepackage{booktabs}
\usepackage{amsmath}
\usepackage{tabularx}
\usepackage{float}

\usepackage[colorlinks=true,breaklinks=true]{hyperref}
\hypersetup{allcolors=[rgb]{0.0 0.0 0.6},linkcolor=[rgb]{0.75 0.05 0.05}}

\usepackage{color}

\usepackage{newtxtext}       %
\usepackage[varvw]{newtxmath}       

\makeindex             

\begin{document}

\title*{New ideas on the formation and astrophysical detection of primordial black holes} 
\author{Marcos M. Flores and Alexander Kusenko}
\institute{Marcos M. Flores \at 
Laboratoire de Physique de l'\'{E}cole Normale Sup\'{e}rieure, ENS, Université PSL, CNRS, Sorbonne Universit\'{e}, Universit\'{e} Paris Cit\'{e}, F-75005 Paris, France
\and Alexander Kusenko \at 
Department of Physics and Astronomy, University of California, Los Angeles, California, 90095-1547, USA\\
Kavli Institute for the Physics and Mathematics of the Universe (WPI), UTIAS, The University of Tokyo, Kashiwa, Chiba 277-8583, Japan 
}
\maketitle

\abstract{
Recently, a number of novel scenarios for primordial black hole (PBH) formation have been discovered.  Some of them require very minimal new physics, some others require no new ingredients besides those already present in commonly considered models, such as supersymmetry.  At the same time, new strategies have emerged for detection of PBHs.  For example, an observation of an orphan kilonova unaccompanied by the gravitational waves signal of merging neutron stars, but associated with a fast radio burst, could be a smoking gun of PBH dark matter.  We review some new ideas for PBH formation and detection. 
}

\section{Introduction} \label{sec:Introduction}

Black holes are known to exist in nature, and can be produced in stellar explosions.  However, the dense early Universe could have provided an addition channel for producing black holes, including supermassive black holes, as well as those with masses much smaller than a solar mass.  The latter is a particularly exciting possibility because in the range of masses $10^{17}-10^{23}$~g, primordial black holes (PBHs) could explain the entirety of dark matter.  The fact that black holes are known to exist makes PBH a very appealing dark matter candidate~\cite{Zeldovich:1967,Hawking:1971ei,Carr:1974nx,Khlopov:1985jw,Dolgov:1992pu,Yokoyama:1995ex,Wright:1995bi,GarciaBellido:1996qt,Kawasaki:1997ju,Green:2004wb,Khlopov:2008qy,Carr:2009jm,Frampton:2010sw,Kawasaki:2016pql,Carr:2016drx,Inomata:2016rbd,Pi:2017gih,Inomata:2017okj,Garcia-Bellido:2017aan,Georg:2017mqk,Inomata:2017vxo,Kocsis:2017yty,Ando:2017veq,Cotner:2016cvr,Cotner:2017tir,Cotner:2018vug,Sasaki:2018dmp,Carr:2018rid,Germani:2018jgr,Banik:2018tyb,Escriva:2019phb,Germani:2019zez,1939PCPS...35..405H,Cotner:2019ykd,Kusenko:2020pcg,deFreitasPacheco:2020wdg,Takhistov:2020vxs,Biagetti:2021eep,Riotto:2024ayo}.

A very popular scenario for PBH formation involves production of relatively large density perturbations on specific small scales~\cite{Zeldovich:1967,Hawking:1971ei,Carr:1974nx,Carr:1975qj,Yokoyama:1995ex,GarciaBellido:1996qt,Kawasaki:1997ju,Kawasaki:2012kn}.  When these perturbations re-enter the horizon, black holes form with masses of the order of the horizon size.  

An alternative class of scenarios involves overdensities on subhorizon scales, which can become black holes with masses much smaller than the horizon at the time of formation.  It was realized relatively recently that many models of particle physics beyond the Standard Model already have all the ingredients for PBH formation.  For example, the widely studied field of supersymmetry (including its minimal version, the Minimal Supersymmetric Standard Model) has PBHs as one of its generic dark matter candidates, thanks to the flat directions predicted within its framework~\cite{Gherghetta:1995dv}.  

Another important and recent development is a new understanding of strategies to search for dark matter in the form of PBHs, including through microlensing observations and using neutron stars as detectors.

We will review some of the recently proposed PBH formation scenarios, as well as new and promising detection techniques. 

\section{PBHs from Yukawa interactions} 

We begin with a remarkably simple scenario which involves only one fermion and one boson in a hypothesized dark sector.  If these two fields interact via a Yukawa coupling, this can potentially lead to PBH formation. 

\subsection{Primordial structure formation}

Black hole formation, in the astrophysical context, arises from the evolution and collapse of heavy stars. Under the standard assumptions of $\Lambda$CDM, the formation of structure only occurs after matter-radiation equality. To see this, note that the matter density perturbations, $\delta_m \equiv (\rho_m - \bar{\rho}_m)/\bar{\rho}_m$, as described by cosmological perturbation theory, evolve via the differential equation
\begin{equation}
\ddot{\delta}_m + 2H\dot{\delta}_m
=
4\pi G_Na^2\bar{\rho}_m\delta_m
\end{equation}
where $\dot{}\equiv d/dt$, $a$ is the scale factor, $H \equiv \dot{a}/a$ is the Hubble parameter and $\bar{\rho}_{m}$ background value of the matter density. In the early, radiation dominated Universe $a\propto t^{1/2}$ and $\bar{\rho}_m\sim 0$ enabling us to conclude that growing mode
\begin{equation}
\delta_m(a) \propto \ln a,
\qquad 
(
\text{gravity only}
)
.
\end{equation}
The standard conclusion of this result is that matter perturbations cannot grow during radiation domination. Naturally, this result implies any non-trivial structure like stars, galaxies, etc., cannot form at these early stages in the evolution of the Universe. Without these objects, the formation black holes in a fashion similar to astrophysical black hole formation appears out of reach.

The inability for structure form during radiation domination is related, in part, to the feeble strength of gravity. Therefore, enabling a phase of \textit{primordial structure formation} requires the introduction of an additional, stronger, force~\cite{Ayaita:2012xm, Casas:2016duf, Amendola:2017xhl, Savastano:2019zpr, Flores:2020drq, Domenech:2021uyx, Domenech:2023afs, Flores:2023zpf}. The natural candidate for such a force is a scalar-mediated Yukawa interaction:
\begin{equation}
\mathcal{L} 
\supset
\frac{1}{2} m_\varphi^2\varphi^2
+
(m_\psi
+
y\varphi)
\bar{\psi}\psi
.
\end{equation}
Scalar-mediated Yukawa interactions lead to forces which are always attractive. Generically, these forces are also always stronger than gravity. This becomes apparent by comparing the couplings of the two forces,
\begin{equation}
\beta
\equiv
y
\left(
\frac{M_{\rm Pl}}{m_\psi}
\right)
\end{equation}
where $M_{\rm Pl}$ is the reduced Planck mass $M_{\rm Pl}^2 = 1/8\pi G_N\sim 2.4\times 10^{18}$ GeV. The appearance of $M_{\rm Pl}$ in the above ratio implies that generically, $\beta\gg 1$. The well-known Yukawa interaction potential,
\begin{equation}
V(r) = -\frac{y^2}{4\pi r}e^{-m_\varphi r}
\end{equation}
illustrates that for length scales $r < m_\varphi^{-1}$, the Yukawa potential is Coulomb-like. 
In the cosmological context, this fact is particularly applicable, especially when the horizon size is smaller than the Compton wavelength of the mediator mass, or equivalently, $m_\varphi < H$.

Before discussing the details of primordial structure formation, we will first specify a set of conditions required before any appreciable structure can form. First, we will require that $\bar{\psi}\psi \leftrightarrow \varphi\varphi$ interactions freeze-out. This is to establish a fixed population of fermions out of which to generate nontrivial structure. Second, we will require that the $\psi$ fermions are non-relativistic. Lastly, we will demand that $\varphi$-radiation pressure is negligible or, in other words, that the scalar mean free path is comparable to the Hubble radius. These conditions are satisfied when the temperature of the fermionic dark sector is $T\sim m_\psi$.

With these conditions in mind, we will proceed by using cosmological perturbation theory to describe the growth of matter perturbations in the presence of a long-range scalar-mediated force. The first thing to note is that scalar forces couple to number density rather than energy density, as is the case with gravity. Therefore it is more useful to examine the \textit{number} density contrast, $\delta_\psi = \delta n_\psi/n_\psi$, as opposed to the energy density contrast. The key equation which describes the growth of overdensities in the $\psi$ fluid is~\cite{Domenech:2023afs},
\begin{equation}
\label{eq:PertDE}
\delta_\psi''
+
\frac{2 + 3x}{2x(1 + x)}
\delta_\psi'
=
\frac{3}{2x(1 + x)}
\delta_\psi f_\psi
\left[
1 + \frac{2\beta^2}{1 + (k/m_\varphi)^{-2}}
\right]
\end{equation}
where $'\equiv d/dx$ with $x = a/a_{\rm eq}$, $a_{\rm eq}$ being the scale factor when $\rho_r = \rho_m + \rho_\psi$ and $f_\psi = \rho_\psi/(\rho_m + \rho_\psi) \leq 1$. For readers familiar with cosmological perturbation theory, this equation is similar to the Meszaros Equation, with an explicit dependence on the scale $k$. During the radiation dominated era, i.e., $x\ll 1$ then
\begin{equation}
\delta_\psi
=
c_1 I_0(\sqrt{6\alpha x})
+
c_2 K_0(\sqrt{6\alpha x}),
\qquad
\alpha = 
f_\psi
\left[
1 + \frac{2\beta^2}{1 + (k/m_\varphi)^{-2}}
\right]
,
\end{equation}
where $I_0$ and $K_0$ are the zeroth-order modified Bessel functions. For $6\alpha x\gg 1$, the growing mode becomes exponential
\begin{equation}
\delta_\psi
\sim
I_0(\sqrt{6\alpha x})
\sim
\frac{e^{\sqrt{6\alpha x}}}{
\sqrt{2\pi\sqrt{6\alpha x}}
}
.
\end{equation}
When the characteristic timescale associated with the growth of perturbations, $\delta_\psi/\dot{\delta}_\psi$, is shorter than the Hubble time, then overdensities will rapidly form. Furthermore, one can see that in the limit that $\beta\to 0$ and $x\ll 1$ then $\delta_\psi \sim {\rm const.}$, as is expected.

Once the perturbations reach $\delta_\psi\sim 1$, the dynamics of structure formation become non-linear and Eq.~\eqref{eq:PertDE} no longer holds. Physically, this corresponds to the formation of virialized $\psi$ halos whose eventual collapse will lead to the formation of black holes. 

The breakdown of linear perturbation theory sets the stage for numerical studies. Before discussing recent numerical work related to primordial structure formation, we will point out that the preceding analysis provides a simplistic picture, which in actuality, could be more complicated. In particular, the fermion $\psi$ acquires and effective mass $m_{\rm eff}(t) = m_\psi + y\varphi(t)$ when $\varphi$ is non-zero. This induces a time-dependent length-scale $\ell(t)$ which significantly alters the simple narrative discussed above. For a comprehensive discussion of the implications of these effects, we point the reader to Refs.~\cite{Domenech:2021uyx, Domenech:2023afs}.

\begin{figure}[h!]
	\centering
	\includegraphics[width=0.95\textwidth]{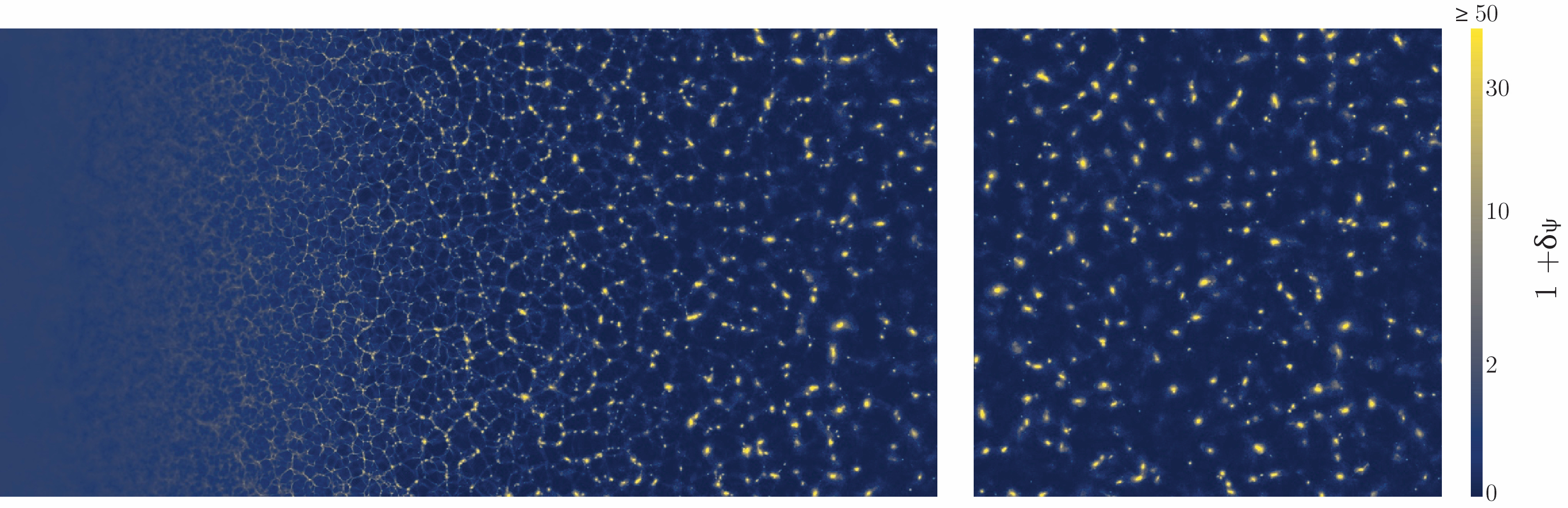}\\
	\caption{Numerical simulations of primordial structure formation. Starting with small perturbations, the above figure demonstrates the growth of nonlinear structure during a radiation dominated era. Adapted from Ref.~\cite{Domenech:2023afs}}
	\label{fig:PSF_Numerical}.
\end{figure}

Figure~\ref{fig:PSF_Numerical} illustrates the growth of structure throughout a radiation dominated era. As is typical in structure formation, halos form as nodes within a network of filaments. However, as time evolves the halos become more concentrated with their maximum radius being determined by the inverse mass of the mediator $\varphi$.

\subsection{Energy dissipation and collapse}

Yet another advantage of the long-range Yukawa interaction is its ability to facilitate dissipation through scalar radiation. Just as accelerating electric charges cause the emission of electromagnetic waves, accelerating fermions, coupled to a scalar mediator, emit scalar waves. Without dissipation, the $\psi$ halos formed via long-range scalar interactions would simply remain as stable virialized halos, assuming the constitute particles are stable~\cite{Savastano:2019zpr}. While this alone is an interesting possibility, dissipation is an unavoidable consequence of Yukawa interactions.

The dynamics of scalar radiation are complicated and heavily dependent on the characteristics of individual halos. For simplicity, we will outline the general history of a given $\psi$ halo. The details of this evolution are an area of active research, and future analytic and numerical work will hopefully shed light on the complex dynamics which can occur as scalar radiation becomes important.

There are various different radiation channels available for scalar emission, which can broadly be classified into oscillatory and non-oscillatory channels. Oscillatory channels are related to the rotation of a given halo. Coherent oscillations correspond to rotation of the entire halo with one single frequency $\omega$. The power radiated through coherent motion is simply related to the coupling $y$ and the number of particles within the halo $N_h$ through $P_{\rm coh} \propto y^2 N_h^2$. Incoherent oscillations correspond to rotation of different particles, or subhalos, with different oscillation frequencies $\omega$. The energy radiated in this case instead scales as $P_{\rm incoh}\propto y^2 N_h$, as is required by the low number density limit.  

The non-oscillatory emission channels involve microscopic phenomena which produce scalar quanta. The first non-oscillatory channel we will discuss is a consequence of pair-wise interactions of fermions, which lead to scalar emission, i.e., bremsstrahlung. This process is similar to free-free emission of photons from plasma~\cite{Maxon:1967, Maxon:1972}. However, unlike the plasmas considered in traditional astrophysical contexts, which contain oppositely charged particles, our system contains only identical particles. This being the case, bremsstrahlung radiation due to two-particle collisions is quadrupole rather than dipole and analogous the $e$-$e$ component of free-free mission from plasma~\cite{Maxon:1967, Maxon:1972}. Similar in spirit to bremsstrahlung is bound state formation, which also leads to the emission of scalar radiation. Bound state formation will be important in the early stages of halo formation, when the interacting particles are non-relativistic and the relative kinetic energy between particles is sufficiently low compared to the binding energy $|E_{\rm bind}| \sim m_\psi y^4$. However, as the halo collapses, high-energy scalars could re-ionize any $\psi$ bound states which may have formed at the early stages of cooling.

The last channel we will discuss deserves special attention because it is the inevitable fate of any collapsing halo. At a given point, scalar radiation will become trapped as the optical depth of $\varphi$ particles decreases within the increasingly dense halo. Once this stage of the evolution is reached, radiation can only escape from a thin surface layer of the collapsing halo. Cooling from the surface will drive less dense, higher energy $\psi$ plasma upward. Simultaneously the cooler, denser, outer layer will fall inward. This drives convection currents which rapidly cool the fireball, leading the halo toward total collapse.

The emission channels discussed above involve complicated, non-linear dynamics which require further investigation to fully understand. Despite the complicated physics involved it is possible to demonstrate that the timescales associated with the processes discussed above are sufficiently short to facilitate efficient collapse. Associated with a given halo is its radiative cooling timescale $\tau_{\rm cool}$ which is defined as
\begin{equation}
\tau_{\rm cool}(R,M_h) 
\equiv
\frac{E}{d E/dt}
=
\frac{E}{P_{\rm brem} + P_{\rm surf} + \cdots}
.
\end{equation}
As indicated above, the cooling rate depends on the halo mass and its radius. In general, the radius is a monochromatically decreasing function of time. For the dissipation channels we discussed above, the cooling rate decreases as $R(t)\to 0$, implying that collapse is a runaway process~\cite{Flores:2020drq}.

Once a given halo has reached a point where its cooling timescale is smaller than the Hubble time $H^{-1}$, the collapse of the halo will commence rapidly on cosmological timescales. The implications of this collapse are wide-ranging~\cite{Flores:2022oef, Durrer:2022cja, Flores:2023nto}, but we will focus only on the possibility of PBH formation.

\subsection{PBH Formation}

The collapse process discussed above will increase the inter-halo particle interaction rate. If the fermions involved are symmetric, this increased number density could lead to particle annihilations that would destroy any structure which may have formed. While this process itself has interesting cosmological implications, it acts as a roadblock for the formation of PBHs. 

To avoid this issue, we will require a particle asymmetry in the fermion sector, i.e.,
\begin{equation}
\eta_\psi 
\equiv
\frac{n_\psi - n_{\bar{\psi}}}{s}
\end{equation}
where $s$ is the comoving entropy number density. Requiring a particle asymmetry is motivated by existing asymmetric dark matter models, e.g.,~\cite{Petraki:2013wwa, Zurek:2013wia}, as well as the existence of the asymmetry in the Standard Model sector. 
The inclusion of an asymmetry in the $\psi$ fermion sector ensures that annihilations will not disrupt the collapse of previously formed $\psi$ halos, and provides a path toward PBH formation.

The last remaining potential roadblock is Fermi degeneracy pressure. Earlier examinations of scalar-mediated fermionic dark matter have demonstrated that larger bound states can form and act as macroscopic dark matter~\cite{Wise:2014jva, Gresham:2017zqi, Gresham:2018rqo, Gresham:2017cvl, Flores:2023zpf}. For the formation of large PBHs Fermi degeneracy pressure can be ignored. To see this, we compare the average distance between $\psi$ fermions within a given halo to the Compton wave length of the fermions. This leads to an expression analogous to the Chandrasekhar limit,
\begin{equation}
M_{\rm PBH}
\geq
\frac{(3/4\pi)^2}{(2G_N)^{3/2}m_\psi^{2}}
\simeq
1.3\times 10^{-3}\ M_\odot\ 
\left(
\frac{5\ {\rm GeV}}{m_\psi}
\right)^2
.
\end{equation}
The above expression tells us that for PBH masses exceeding $\sim 10^{-3}\ M_\odot$, Fermi degeneracy pressure never comes into play. For smaller masses, one must consider the fermionic pressure which may inhibit the formation of a PBH. In order to ensure that Fermi degeneracy pressure is not an issue, we require that the Fermi energy be small compared to the potential energy $y^2 N/R$ as $R \to R_S$. This condition is similar to the Chandrasekhar limit but is modified by a factor of $\beta^{-3/2}$. For the parameters relevant for PBH formation, this condition will not hamper the formation of PBHs relevant for dark matter~\cite{Flores:2020drq}. 

The abundance of PBHs will be determined by the initial number density of the $\psi$ fluid. Given that we are considering asymmetric fermions, the number density is simply given by $\eta_\psi T^3$. Under the assumption that all of the fermions end up in halos and therefore fall into PBHs, the PBH abundance $f_{\rm PBH}$ will be given by
\begin{equation}
\label{eq:PSF_fPBH}
f_{\rm PBH}
\equiv
\frac{\Omega_{\rm PBH}}{\Omega_{\rm DM}}
=
0.2
\frac{m_\psi}{m_p}\frac{\eta_\psi}{\eta_{\rm B}}
=
\left(
\frac{m_\psi}{5\ {\rm GeV}}
\right)
\left(
\frac{\eta_\psi}{10^{-10}}
\right)
.
\end{equation}
Another advantage of using an asymmetric fermion to generate PBHs comes directly from the above formula. In particular, this scenario inherits the motivating feature behind asymmetric dark matter, namely an explanation for the factor five difference between the dark matter and visible matter energy densities. 

The mass distribution of PBHs generated via primordial structure formation will be an extended distribution. An absolute lower limit can be placed on the minimum size of a PBH formed via this mechanism. In particular, the Compton wavelength of the $\psi$ particles must fit within the Schwarzschild radius of their prospective PBH. This gives
\begin{equation}
M_{\rm PBH}^{\rm min}
\geq \frac{4\pi M_{\rm Pl}^2}{m_\psi}
.
\end{equation}
The maximum PBH size is determined by the total $\psi$ mass within the Hubble volume at the time of formation. A precise understanding of the mass distribution of $\psi$ halos, and therefore PBHs, has yet to be established. Following Ref.~\cite{Flores:2020drq}, we will adopt a distribution motivated by the Press–Schechter formalism. In particular, we will assume that the halo mass distribution is described by
\begin{equation}
M^2
\frac{dN_h}{dM}
\propto
\frac{1}{\sqrt{\pi}}
\left(
\frac{M}{M_*}
\right)^{1/2}
e^{-M/M_*}
\end{equation}
where the overall normalization is set by requiring that the integrated mass function gives $f_{\rm PBH}$ as determined by Eq.~\eqref{eq:PSF_fPBH}. For details about the precise nature of the mass function, we point the reader to Ref.~\cite{Flores:2020drq}.

\begin{figure}[htb]
    \centering
	\includegraphics[width=0.8\linewidth]{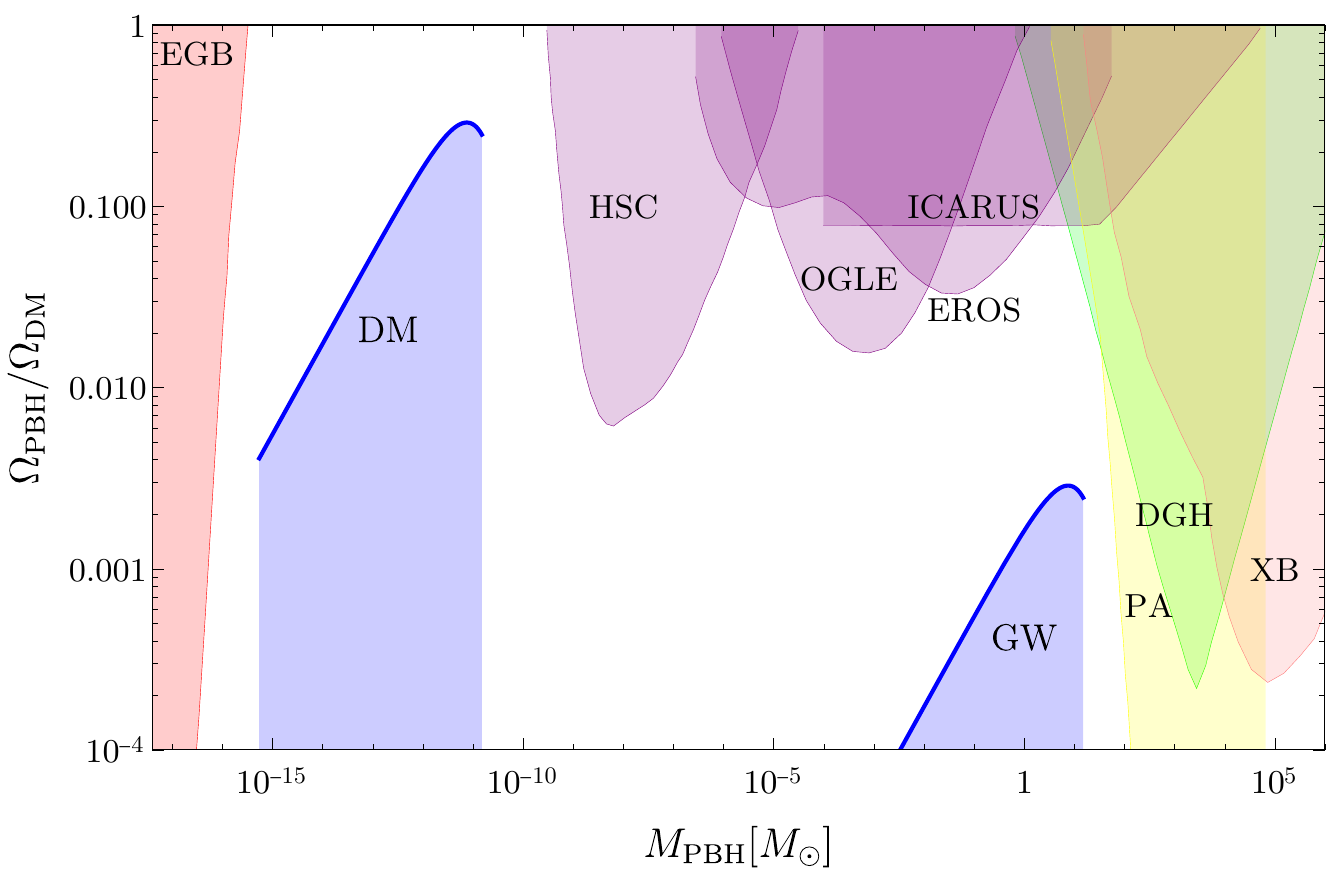}\\
	\caption{The PBH abundance within the primordial structure formation framework. The dark matter (DM) curve corresponds to $\eta_\psi\sim \eta_B\sim 10^{-10}$ with $m_\psi = 5$ GeV. Alternatively, the gravitational wave curve (GW) relevant to LIGO, Virgo, and KAGRA, corresponds to $\eta_\psi = 10^{-9}$ and $m_\psi = 5$ MeV. See Ref.~\ref{fig:PSF_fPBH_fig} for further details. The constraints plotted are reproduced from Refs.~\cite{Carr:2020gox, Carr:2020xqk} and the references within.}
	\label{fig:PSF_fPBH_fig}
\end{figure}

Figure~\ref{fig:PSF_fPBH_fig} illustrates the mass function for two set of parameters. Taking $m_\psi = 5$ GeV and $\eta_\psi = 10^{-10}$ (which is comparable to the observed value of the baryon asymmetry) gives $f_{\rm PBH} = 1$ and produces a distribution which sits within the unconstrained dark matter window. Alternatively, one can adjust the parameters to $\eta_\psi = 10^{-9}$ and $m_\psi = 5$ MeV which leads to a population of PBHs which would act as a subcomponent of dark matter, but might be relevant for present-day gravitational wave experiments.

\subsection{Observational signatures}

Primordial structure formation in general provides a rich phenomenology with many possible observable signals. For example, the introduction of a light degree of freedom though the scalar $\varphi$ could potentially modify $\Delta N_{\rm eff}$. Not only is this allowed, but it may be favorable in addressing the so-called Hubble tension~\cite{Bernal:2016gxb,Gelmini:2019deq,Anchordoqui:2019yzc,Vattis:2019efj,Escudero:2019gvw,Gelmini:2020ekg,Vagnozzi:2019ezj,Wong:2019kwg, Perivolaropoulos:2021jda}. Apart from this, two particular observables are of interest in the context of PBHs, namely the expected spin distribution and the emission of gravitational waves.

\subsubsection{Spin distribution}

An advantage of using primordial structure formation and scalar radiation for the generation of PBHs is the efficient removal of both energy and angular momentum. In particular, the efficient removal of angular momentum avoids the need to generate spherical configurations of matter which are required for PBH formation in the gravity-only context. 

The removal of angular momentum of a spinning fermion halo is due to two possible effects. First, rotation of the halo as a whole will lead to emission of energy, which is directly related to the rate of angular momentum loss. Specifically, if decomposed in a spherical basis one can demonstrate that,~\cite{Flores:2021tmc}
\begin{equation}
\frac{dJ_{\ell m}}{dt}
=
\frac{m}{\omega}
\left(
\frac{dE_{\ell m}}{dt}
\right)
\end{equation}
where $\ell$ and $m = -\ell, \ldots, \ell$ are the spherical expansion indices and $\omega$ is the oscillation frequency. Alternatively, one can examine the emission of scalar quanta due to pair-wise scattering. In that case, the loss of angular momentum can be seen to be driven from blueshifting or redshifting of radiation as it is being emitted in a direction parallel or antiparallel relative to the tangential velocity of the rotating halo. This leads to an angular momentum loss  rate which also proportional to the energy loss, but also depends on the tangential velocity in a nontrivial fashion,~\cite{Flores:2021tmc}
\begin{equation}
\frac{dJ}{dt}
=
-
R
\frac{dE}{dt}f(v),
\qquad
f(v)
=
\frac{3\pi^2}{4}
\left[
\frac{(1+v^2)\tanh^{-1}v - v}{v^2}
\right]
\end{equation}
where the function $f(v)$ accounts for the red(blue)shifting of the scalar quanta. For small velocities $f(v\ll 1)\propto v$.

As demonstrated in Ref.~\cite{Flores:2021tmc}, the characteristic timescale associated with angular momentum loss $\tau_J \equiv J/(dJ/dt)$ is smaller than the cooling timescale, $\tau_{\rm cool}$, for the parameters relevant for PBH formation. This implies that angular momentum losses are rapid, and any halo which may have formed initially spinning will rapidly shed its angular momentum. Given that this is the case, any PBH which may have formed through primordial structure formation and collapse should be spinless at the time of formation. Naturally, accretion and mergers can alter the PBH spin distribution after formation. We will leave discussion of these effects to Ref.~\cite{DeLuca:2023bcr}.
 
\subsubsection{Gravitational waves}

The key ingredient for the generation of gravitational waves is a time-changing mass quadrupole moment. Given that the $\psi$ perturbations will, in general, collapse in a nonspherical fashion it is natural to expect that gravitational waves will be emitted. Since conventional wisdom requires a matter dominated era to facilitate collapse, early works such as Refs.~\cite{Assadullahi:2009nf, Jedamzik:2010hq, Dalianis:2020gup, Nakama:2020kdc} have previously examined gravitational waves in this context. The generation of gravitational waves from isocurvature perturbations in a radiation dominated era have also been examined in the literature~\cite{Ananda:2006af, Baumann:2007zm, DeLuca:2019llr, Domenech:2021and}. The majority of the literature referenced uses cosmological perturbation theory to determine the gravitational wave signal. While this should also be possible within the framework of primordial structure formation, the details are nontrivial. The one difficulty arises from the fact that the scalar mediator couples to the fermionic number density as opposed to the energy density. Analysis of the growth of perturbations, including the discussion above, is always performed in the Newtonian limit. Given that the generation of gravitational waves is a relativistic effect, more work needs to be done to describe the growth of overdensities in a fully relativistic framework.

To avoid these issues, Ref.~\cite{Flores:2022uzt} generalized the framework of Ref.~\cite{Dalianis:2020gup} which utilizes the Zeldovich  approximation~\cite{Zeldovich:1970gia}. The Zeldovich approximation allows us to examine the time-changing mass quadrupole moment of individual matter perturbations. By doing so, we can avoid using cosmological perturbation theory and also examine dynamics beyond the traditional regime permissible in a linear framework. 

\begin{figure}[htb]
	\centering
	\includegraphics[width=0.95\linewidth]{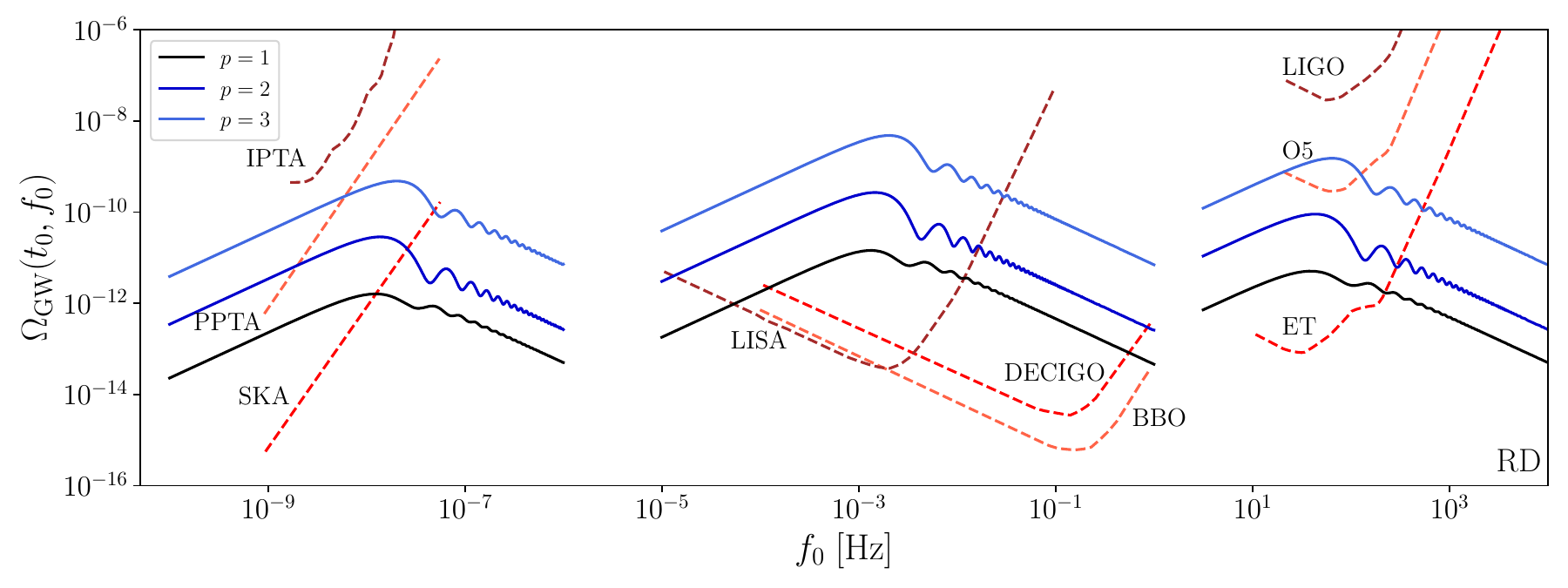}\\
	\caption{Predicted gravitational wave signals due to the formation of structure due to a long-range interaction. The distribution of masses of the overdensities are assumed to be monochromatic. Here, the masses of the overdensities are  $10^{-2}$ $M_\odot$ (left), $10^{-12}$ $M_\odot$ (center), and $10^{-21}$ $M_\odot$ (right). For further details, see~\cite{Flores:2022uzt} and the references within.}
	\label{fig:PSF_GWs}
\end{figure}

The details of the calculations involved are quite complicated and are presented in Refs.~\cite{Harada:2016mhb, Dalianis:2020gup, Flores:2022oef}. To parameterize the growth of additional structure, Ref.~\cite{Flores:2022uzt} assumed that fermion perturbations grow as
\begin{equation}
\delta_\psi 
\propto
a^p
\end{equation}
where $p$ parametrizes the strength of long-range scalar interaction. In Fig.~\ref{fig:PSF_GWs}, we present the signal of gravitational waves for a broad range of frequencies. The peak of the distribution is determined by the time in which the $\delta_\psi$ enters into the horizon. This time is associated with the mass of the perturbation, with $f_{\rm max}\propto M^{-1/2}$. It should also be noted that the signals presented in Fig.~\ref{fig:PSF_GWs} assume a monochromatic distribution of overdensities.

Outside of the context of PBHs, gravitational waves produced through primordial structure formation offer an opportunity to examine presence of long-range interactions in the early Universe. This effect, combined with traditional particle physics detection methods, offers a unique, multi-approached strategy for probing physics beyond the Standard Model.

\subsection{Future directions}

Primordial structure formation remains an active field of research which requires multidisciplinary collaboration to tackle many of the remaining outstanding questions. Numerical simulations will continue to probe the formation of structure during radiation domination. At the present moment, these simulations do not include dissipation due to scalar radiation. Dissipation could dramatically impact the mass distribution of halos formed, and is an important factor for PBH formation. Numerical simulations will also be important in understanding the later stages of collapse, particularly during the surface cooling phase. These open questions, and others, continue to propel our exploration of the early Universe toward new and exciting directions.

\section{PBHs as a generic dark matter candidate in MSSM and other models with supersymmetry}
It is important to realize that PBH production is a fairly generic consequence of supersymmetry.  For example, the Minimal Supersymmetric Standard Model (MSSM) has all the ingredients necessary for PBH production.  Whether or not PBH are produced and the abundance of PBH depend on the details of supersymmetry breaking.  

\subsection{Flat directions, Q-balls, and PBHs}
What makes the production of PBHs possible is the existence of flat directions in the MSSM and other supersymmetric generalizations of the Standard Model.  These flat directions $\phi$ correspond to zeros of the scalar potential: $U(\phi)=0$ in the limit of unbroken supersymmetry~\cite{Gherghetta:1995dv}.  Supersymmetry breaking lifts the flat direction in a model-dependent way, but the lifted flat directions remain D-flat, so that they grow logarithmically or at most quadratically.   Such directions in the scalar potential develop large vacuum expectation values (VEV) in the early Universe.  This is due to two reasons.   

First, the effective minimum of the potential in de Sitter space can be displaced because of the terms $\sim c H^2 \phi^2$ that come from the Kaehler potential~\cite{Affleck:1984fy,Dine:1995uk}.  If $c<0$, the field acquires a nonzero VEV. 

Second, regardless of where the minimum is, the field does not remain in that minimum everywhere in space during de Sitter expansion.  Instead, the field undergoes fluctuations on superhorizon scales~\cite{Chernikov:1968zm,Tagirov:1972vv,Bunch:1978yq,Linde:1982uu,Lee:1987qc,Starobinsky:1994bd}. The results of these papers can be summarized as follows: in de Sitter space, each scalar field is displaced from the minimum, and each degree of freedom, on average, carries the amount of energy density of the order of $H_{\rm infl}^4$, where $H_{\rm infl}$ is the Hubble parameter during inflation. (Since every degree of freedom carries the average energy $\sim H_{\rm infl}$, one can think of $H_{\rm infl}$ as the effective temperature of de Sitter space for the purposes of scalar fluctuations.) 
For some simple potentials, the shift in the minimum due to first effect  and the average deviation from that minimum due to second effect can be of the same order of magnitude.  

An initially homogeneous condensate can fragment into stable or relatively stable Q-balls~\cite{Kusenko:1997si}, which become the building blocks of PBHs.  The population of Q-balls behaves as matter, but it is matter with relatively few particles per horizon.  Therefore, the Poisson fluctuations are unsuppressed, and some fluctuations can be big enough to collapse to black holes~\cite{Cotner:2016cvr,Cotner:2017tir,Cotner:2019ykd}.  

Supersymmetry provides a natural framework for PBH formation from scalar field fragmentation.  It is particularly interesting that electroweak-scale supersymmetry naturally results in the PBH masses for which there are no observational constraints and which can account for all dark matter. While Q-ball formation is possible in different models of supersymmetry breaking, let us consider gauge-mediated supersymmetry breaking at the scale $M_{\rm SUSY}\sim 100 \ {\rm TeV}$.  At the time of fragmentation,  when the energy density of the condensate is a fraction  $f\sim r_f^{-1/2}\lesssim 1$ of the total energy density ($\rho_\phi \sim M_{\rm SUSY}^4 \equiv f \times \rho_{\rm tot}$), the mass inside the horizon is 
\begin{equation}
M_{\rm hor}\sim r_f^{-1/4} 
\left(\frac{M_{\rm Pl}^3}{M_{\rm SUSY}^2}\right)\sim 10^{22} {\rm g} 
\left ( \frac{100~ {\rm TeV}}{M_{\rm SUSY} }\right)^2~. 
\end{equation}
Let the number of Q-balls per the horizon be $N_H\sim 100$. A typical PBH results from coalescence of 10 to 100 Q-balls, and the mass of a typical PBH is
\begin{equation}
    M_{\rm PBH} \sim r_f^{-1/4} \times  10^{22} {\rm g} \left ( \frac{100~ {\rm TeV}}{M_{\rm SUSY} }\right)^2~,  
\end{equation}
which is consistent with the open window for dark matter in the form of PBH at masses 
\begin{equation}
    10^{17} {\rm g} \lesssim M_{\rm PBH} \lesssim 10^{22} {\rm g}~. 
\end{equation}

It is intriguing that supersymmetry just above the electroweak scale predicts the masses of PBHs consistent with the current bounds for dark matter.

\subsection{Long-range interactions mediated by flat directions}

The existence of flat directions allows for another PBH formation scenario, which also starts with fragmentation of the scalar condensate into Q-balls but then proceeds in a different dynamical fashion.  The previously mentioned scenario requires the introduction of an intermediate matter dominated era to facilitate the collapse of overdensities via gravity. This requirement is again a consequence of the fact that traditionally, matter perturbations only grow during a matter dominated phase.  Given that gravity is relatively inefficient in removing energy and angular momentum, only rare, spherical configurations can ultimately collapse into black holes. This limitation led the authors of Ref.~\cite{Flores:2021jas} to pursue a less limited framework.

With the framework of primordial structure formation in mind, Ref.~\cite{Flores:2021jas} included an additional scalar force which mediates interactions between Q-balls. Just as in the case of primordial structure formation, this scalar interaction can lead to the coalescence and collapse of overdensities, even during radiation domination. As discussed before, the removal of energy and angular momentum is very efficient for scalar waves, thus alleviating the need for special spherical configurations of particles.

As with primordial structure formation, the dynamics in this scenario are complicated, and numerical simulations will play a key role in future investigations. However, given the size and strengths of the forces involved, the expectation is that after scalar fragmentation large systems of charges will be drawn together. Ultimately, these charges will start to collide and merge, resulting in a population of larger and larger charges. For Q-balls formed from the fragmentation of flat directions, their mass and radius is dictated by,~\cite{Dine:2003ax}
\begin{equation}
M_Q = M_{\rm SUSY} Q^{3/4},
\qquad
R_Q = \frac{Q^{1/4}}{M_{\rm SUSY}}
\end{equation}
where $Q$ is the $U(1)$ charge associated with the flat direction which permits the formation of Q balls. Setting $R_Q = R_S = 2G_NM_Q$ establishes threshold where a charge will eventually fit within its own Schwarzschild radius. This leads to a predicted PBH mass which depends only on the SUSY breaking scale,
\begin{equation}
M_{\rm PBH}
\sim 10^{22}\ {\rm g}\ 
\left(
\frac{100\ {\rm TeV}}{M_{\rm SUSY}}
\right)^2
.
\end{equation}
Compared to the previous, gravity-only mechanism, the mass dependence is nearly exactly the same. The formation of PBHs via supersymmetry and long-range scalar forces has the ability to produce populations of PBHs which can be the totality of dark matter, or which may explain microlensing observations by HSC and OGLE~\cite{Niikura:2017zjd,Niikura:2019kqi,Sugiyama:2021xqg}.

\subsubsection{Tests for high-scale supersymmetry}

The fact that the PBH mass is inversely proportional to the scale of supersymmetry breaking leads to the opportunity to examine high-scale supersymmetry by studying the implications of light PBHs. Although PBHs with masses below $\lesssim 10^{17}$ g would have evaporated by the present-day, the implications of a light population of PBHs could still be observable today. In particular, if these light PBHs had the opportunity to dominate the energy density of the Universe, than a detectable gravitational wave signal can be produced via the \textit{poltergeist mechanism}~\cite{Inomata:2019ivs, Inomata:2019zqy, Domenech:2020ssp, Papanikolaou:2020qtd, Bhaumik:2020dor, Inomata:2020lmk, White:2021hwi, Domenech:2021wkk, Domenech:2023mqk, Harigaya:2023mhl, Flores:2023dgp, Pearce:2023kxp}. 

The poltergeist phenomenon is a result of the fact that curvature perturbations source gravitational waves at second order in perturbation theory~\cite{Assadullahi:2009nf, Alabidi:2013lya, Kohri:2018awv}. Two features of an intermediate dominated era help in generating a larger gravitational wave signal. First, unlike during radiation domination, the gravitational potential during a matter dominated era does not decay, even on subhorizon scales. Second, if the transition from intermediate matter domination to radiation domination is rapid, then there is a resonant enhancement.  This phenomenon arises from the quick conversion of matter overdensities into relativistic sound waves, and the enhancement of the comoving gravitational waves that travel with these sound waves. 

The strength of the gravitational wave signal produced from an intermediate matter dominated era and PBH evaporation depends on the length of the matter dominated era, and the speed of the transition from matter domination to radiation domination. Within the framework for supersymmetric PBH formation we have discussed, there exists a large parameter space which could generate PBHs with the masses required to produce the necessary conditions for generating a detectable gravitational wave signal. As highlighted in Ref.~\cite{Flores:2023dgp}, a wide range of supersymmetric breaking scales can be explored with many of proposed future gravitational wave detectors.

\section{Neutron stars as a PBH detector}

If PBHs comprise dark matter, their number density is very small for the allowed mass window.  Indeed, no more than one dark-matter black hole could be found in the vicinity of the solar system, and that black hole with an asteroid-scale mass would have a Schwarzschild radius which is smaller than the wavelength of visible light.  If such a black hole passes through a star or a planet, the drag forces are not sufficient to stop the black hole. However, PBHs can get captured by neutron stars, thanks to their higher density. This will occur in regions with high dark matter densities, like the galactic center. Once a PBH settles inside the neutron star, it consumes the neutron star from the inside, turning it into a 1--2 $M_\odot$ black hole~\cite{Kouvaris:2013kra,Fuller:2017uyd, Takhistov:2017nmt, Takhistov:2017bpt, Takhistov:2020vxs}.
This offers an exciting possibility to use neutron stars as a PBH detectors.  The evidence of a neutron star disruption by a PBH consists of several observable effects: a kilonova explosion~\cite{Fuller:2017uyd}, a fast radio burst~\cite{Fuller:2017uyd,Abramowicz:2017zbp}, and a population of 1--2 $M_\odot$ black holes which can be detected by LIGO/VIRGO, KAGRA, etc.~\cite{Fuller:2017uyd, Takhistov:2017nmt, Takhistov:2017bpt,  Takhistov:2020vxs}

The average life-time of a neutron star in a PBH rich environment is determined by three physical processes. The first, and most important is the PBH capture rate
\begin{equation}
F \equiv
f_{\rm PBH}\times
F_0^{\rm MW}
\end{equation}
where $F_0^{\rm MW}$ is the Milky Way capture rate given by,~\cite{Capela:2013yf}
\begin{equation}
F_0^{\rm MW}
=
\sqrt{6\pi}
\frac{\rho_{\rm DM}}{M_{\rm PBH}\bar{v}}
\left(
\frac{2GM_{\rm NS}R_{\rm NS}}{1 - 2GM_{\rm NS}/R_{\rm NS}}
\right)
\left(
1 - e^{-3E_{\rm loss}/(M_{\rm PBH}\bar{v}^2)}
\right)
\end{equation}
where $\bar{v}$ is the dark matter velocity dispersion, $\rho_{\rm DM}$ is the dark matter energy density and
\begin{equation}
E_{\rm loss}
\approx
58.8\frac{G^2 M_{\rm PBH}^2 M_{\rm NS}}{R_{\rm NS}^2}
\end{equation}
is the average interaction loss energy during a neutron star-PBH interaction. The capture rate may also be enhanced by considering the neutron star velocity dispersion~\cite{Fuller:2017uyd}. However, for the parameter space we consider this effect will be small. For our analysis, we consider neutron stars with masses $M_{\rm NS} = 1.5\ M_\odot$ and radii $R_{\rm NS} = 10$ km. 

Once a PBH is gravitationally bound to a given neutron star, some time is required for the microscopic black hole to settle into its host neutron star's core,~\cite{Capela:2013yf}
\begin{equation}
t_{\rm set}
\simeq 1.3\times 10^{9}\ \text{yr}\ 
\left(
\frac{M_{\rm PBH}}{10^{19}\ {\rm g}}
\right)^{-3/2}
.
\end{equation}
Once settled at the neutron star core, the sublunar PBH will begin to consume the surrounding neutron rich material. Therefore, the final contribution to the average neutron star lifetime is an estimation of the accretion rate. For simplicity, we will assume spherical Bondi accretion,
\begin{equation}
\dot{M}_{\rm PBH} = 4\pi\lambda_s G^2M_{\rm PBH}^2\rho_{\rm NS}/v_s^3,
\end{equation}
where $M_{\rm PBH}$ is the time evolving mass of the growing black hole, $v_s$ is the sound speed, $\rho_{\rm NS}$ is the neutron star density, and $\lambda_s$ is an order one parameter depending on the neutron star equation of state. For a neutron star described by an $n = 3$ polytrope, $v_s = 0.17$, $\rho_{\rm NS} = 10^{15}$ g cm$^{-3}$, and $\lambda_s = 0.707$~\cite{Shapiro:2008CO}, we find
\begin{equation}
t_{\rm acc}
\equiv
\frac{M_{\rm PBH}}{dM_{\rm PBH}/dt}
\simeq
10\ {\rm yr}
\left(
\frac{10^{19}\ {\rm g}}{M_{\rm PBH}}
\right)
.
\end{equation}
Though we are examining incredibly small black holes, the traditional Bondi accretion rate is still reliable for the PBHs in the dark matter mass range~\cite{Giffin:2021kgb}.

The average neutron star lifetime will be given by
\begin{equation}
\langle
t_{\rm NS}
\rangle
=
F^{-1}
+
t_{\rm set}
+
t_{\rm acc}
.
\end{equation}
For the majority of parameter space, the interaction timescale $F^{-1}$ is the main contribution to $\langle t_{\rm NS} \rangle$. Assuming the capture and conversion of PBHs follows a Poisson process, the number of converted neutron stars as a function of time is
\begin{equation}
\label{eq:NSConvFraction}
\Upsilon(t)
=
1 - \exp(-t/\langle t_{\rm NS} \rangle)
.
\end{equation}
For the neutron star parameters mentioned above and $f_{\rm PBH} = 1$, PBHs with masses between $10^{17}$ to $10^{25}$ g lead to an average lifetime of $\langle t_{\rm NS} \rangle \lesssim 10^{12}$ yr~\cite{Fuller:2017uyd}. Such a lifetime suggests that $\mathcal{O}(1-10) \%$ of pulsars in a PBH-rich environment will be consumed within the lifetime of the galaxy. This is consistent with observations which suggest an underabundance of millisecond pulsars within the regions closest to the Milky Way's central supermassive black hole, Sagittarius A$^*$ (Sgr A$^*$)~\cite{Dexter:2014GCPulsProb}.

In general, the destruction of a neutron star is a dramatic event~\cite{Fuller:2017uyd}. If the neutron star hosting a black hole is a rapidly rotating millisecond pulsar, the equatorial regions move close to the speed of light.  During the last milliseconds of the neutron star's demise, its radius decreases dramatically, and, by angular momentum conservation, the star spins up, leading to release of neutron rich material.  This produces a kilonova explosion with similar optical properties as the merger of neutron stars. 

Kilonova explosions are a likely site of the r-process nucleosynthesis.  The only kilonova observed so far was a neutron star merger detected by LIGO.  This detection triggered quick follow-up optical observations, which were generally consistent with a r-process nucleosynthesis site.  However, the rates of neutron star mergers are not sufficient to explain all r-process nucleosynthesis by neutron star collisions~\cite{Kobayashi:2020jes}.  A disruption of a neutron star by PBH would provide an additional contribution to synthesis of heavy elements. 

When a PBH settles in the center of a neutron star and consumes it from the inside, the system does not generate the gravitational waves signal similar to the neutron star merger.  In fact, due to a approximate axial symmetry, any gravitational waves radiation is relatively weak.  In the past, observation of a kilonova event without a signal from LIGO was incredibly improbable.  However, the advent of Rubin LSST, ZTF and future surveys open the opportunity to detect tens of kilonovae per year without the need for a gravitational waves trigger.   
An additional signature of PBH induced kilonova is a fast radio burst~\cite{Fuller:2017uyd,Abramowicz:2017zbp}. During the last milliseconds of the neutron stars collapse, its magnetic field changes rapidly, generating a strong and observable radio signal~\cite{Fuller:2017uyd,Abramowicz:2017zbp}. 

Therefore, observation of an orphan kilonova unaccompanied by the gravitational waves signature of a neutron star merger, but accompanied by a fast radio burst would be a smoking-gun indicator of PBH-induced destruction of a neutron star~\cite{Fuller:2017uyd}. 

\subsection{\textit{G} Objects}

Neutron star destruction via PBHs also offers the opportunity to explain another peculiarity in the galactic center, namely the presence of mysterious class of objects called $G$ objects~\cite{Ciurlo:2020GObjects}. Located within the central inner most region of the galactic center, six $G$ objects have been identified. Dynamically, $G$ objects act as stellar mass objects. Thermally however, these same objects appear to radiate in a similar fashion to dust and ionized gas. 

The two most studied $G$ objects, named G1 and G2, have also demonstrated
a surprising resilience to tidal disruption. When passing close to Sgr A${}^*$ both G1 and G2 remained intact, despite the fact that both objects experienced large tidal forces during their passage. This observation seems to indicate that $G$ objects are not simply clouds of gas, but instead contain a stellar-mass core which provides gravitational stability. Many attempts have been made to explain the nature of $G$ objects. Most models have postulated that the central stellar-mass core is a star. For example, the central star may be a young, low-mass star
that has retained a protoplanetary disk~\cite{Murray:2012Proto} or that generated a mass-loss envelope~\cite{Scoville:2013Mass-LossEnv}. Alternative models of $G$ object formation have viewed the $G$ objects as a merger product of a binary system~\cite{Prodan:2015Secular, Stephan:2016BinaryGOs, Stephan:2019fbg}.

In Ref.~\cite{Flores:2023lll}, the authors offered an origin for $G$ objects which relies on the interactions between PBHs and neutron stars. As indicated before, the destruction of neutron stars will occur primarily in the inner most regions of the galaxy. Furthermore, as stated before, the final product of PBH-induced neutron star destruction will be stellar-mass black hole surrounded by an atmosphere of dust and gas. These objects can be interpreted as the observed population of $G$ objects. As discussed in Ref.~\cite{Flores:2023lll}, the number of expected converted neutron stars are consistent with the number of observed $G$ objects. Additionally, the luminosity of a solar mass black hole cloaked by an atmosphere are consistent with the luminosities inferred from $G$ object observations. These two coincidences point to the possibility that neutron star destruction via sub-lunar PBHs may account for the appearance of these mysterious objects within our galaxy. It is through this lens that the existence of G object-like systems provide another pathway for testing PBHs as a dark matter candidate.

\section{Other scenarios}

The formation of PBHs has continued to remain a topic of much interest within the literature. There is a vast and ever-growing set of models which can accommodate the generation of black holes at the earliest stages of the history of the Universe. Given the enormity of available models, we will be unable to mention all of the possibilities. However, in order to provide the reader with an overview of the field we will include a broad summary of the landscape of non-inflationary PBH formation mechanisms.

\textit{PBHs from phase transitions:} Phase transitions have long-since been considered a possible formation channel for PBHs. A variety of physical phenomena associated with phase transitions can lead to the formation of black holes. Early work considered the collision of bubble walls~\cite{Hawking:1982ga} or the in-fall of the domain wall which separates true- from false-vacuum regions~\cite{Sato:1981gv, Kodama:1982sf, Blau:1986cw}. The latter scenario has the unique feature of forming worm-hole solutions which appear as PBHs to an outside observer, while internal observers would perceive themselves to be in a separate and expanding Universe. This possibility has also been examined within the context of inflation, particularly with vacuum bubbles which may have nucleated during the inflationary phase of the early Universe~\cite{Garriga:2015fdk, Deng:2017uwc, Kusenko:2020pcg}. Recently, there has also been considerable interest in PBHs from supercooled phase transitions~\cite{Sato:1981gv, Kodama:1982sf, Blau:1986cw, Liu:2021svg, Baker:2021sno, Kawana:2022olo, Gouttenoire:2023naa, Baldes:2023rqv, Jinno:2023vnr, Lewicki:2023ioy, Lewicki:2024ghw, Flores:2024lng, Conaci:2024tlc}.

Particles which are massive in the true vacuum state could become trapped in the false vacuum region. These heavy particles would not have the energy to penetrate the bubble walls separating the two regions, and be compressed into a black hole~\cite{Baker:2021sno, Baker:2021nyl, Kawana:2021tde, Kawana:2022lba}.

Phase transitions, like the QCD phase transition, are associated with large changes in the effective number of degrees of freedom. As such, phase transitions have the ability to depress the sound speed. Since the critical collapse criteria in the canonical inflationary scenario is related to the sound speed, PBH production may be enhanced at the scales associated with a given phase transition~\cite{Crawford:1982yz, Jedamzik:1996mr, Carr:2019kxo}. 

\textit{PBHs from topological defects:} Alongside phase transitions, the formation of topological defects, which act as the building blocks of PBHs, may also lead to the formation of PBHs. This includes the collapse of cosmic string loops~\cite{Hawking:1987bn, Polnarev:1988dh, Garriga:1993gj, Caldwell:1995fu, MacGibbon:1997pu, Jenkins:2020ctp}, or the collapse of domain walls~\cite{Rubin:2001yw, Gelmini:2022nim, Gelmini:2023ngs}. 

\textit{Other scenarios and future directions:} Besides those based on phase transitions and topological defects, there are also many other possible PBH formation mechanisms, for example, those based on quark confinement~\cite{Dvali:2021byy}, formation from isocurvature perturbations~\cite{Passaglia:2021jla} or PBH formation from stochastic tunnelling~\cite{Animali:2022otk}.

\textit{Short-lived black holes and supermassive black holes:}
The James Webb Space Telescope allowed for the discovery of supermassive black holes at very high redshifts, when the Universe was less than a billion years old~\cite{Bunker:2023lzn}.  
Small black holes with lifetimes too short to be dark matter could form and decay during the dark ages, after recombination.  Evaporation of such black holes could provide the source of heat necessary for direct collapse of large gas clouds to supermassve black holes~\cite{Lu:2023xoi}.  

Recently, there has been interest in {\it nonprimordial} light black holes forming in the present epoch~\cite{Lu:2022jnp, Picker:2023lup, Picker:2023ybp, Korwar:2024ofe}.  In particular, small black holes forming at present time in the galactic center could simultaneously explain the GeV excess, as well as the flux of cosmic-ray antiprotons and in a few tentative antihelium events reported by the anti-matter spectrometer AMS-02~\cite{Picker:2023lup,Korwar:2024ofe}.

\begin{acknowledgement}
The work of M.M.F was supported by the European Union’s Horizon 2020 research and innovation programme under grant agreement No 101002846, ERC CoG CosmoChart.
A.K. was supported by the U.S. Department of Energy (DOE) Grant No. DE-SC0009937; by World Premier International Research Center Initiative (WPI), MEXT, Japan; and by Japan Society for the Promotion of Science (JSPS) KAKENHI Grant No. JP20H05853.
\end{acknowledgement}

\bibliographystyle{unsrt}
\bibliography{Chapter_new_production}

\end{document}